\documentclass[fleqn,usenatbib]{mnras}

\usepackage{newtxtext,newtxmath}

\usepackage[T1]{fontenc}
\usepackage{multirow} 
\usepackage{ragged2e}
\DeclareRobustCommand{\VAN}[3]{#2}
\let\VANthebibliography\thebibliography
\def\thebibliography{\DeclareRobustCommand{\VAN}[3]{##3}\VANthebibliography}

\usepackage{graphicx}	
\usepackage{amsmath}	
\graphicspath{{./}{plots/}}
\usepackage{tabularx}
\usepackage{caption}

\title{Diagnosing the Properties and Evolutionary Fates of Black Hole and Wolf-Rayet X-ray Binaries as Potential Gravitational Wave Sources for the LIGO-Virgo-KAGRA Network}
\title[Diagnosing BH-WR X-ray binaries as Potential GW Sources for the LVK Network]{Diagnosing the Properties and Evolutionary Fates of Black Hole and Wolf-Rayet X-ray Binaries as Potential Gravitational Wave Sources for the LIGO-Virgo-KAGRA Network}
\author[Zi-Yuan\, Wang et al.]{Zi-Yuan\, Wang,$^{1}$
Ying\, Qin,$^{1}$\thanks{E-mail: yingqin2013@hotmail.com}
Georges\, Meynet,$^{2,3}$
Qing-Zhong\, Liu,$^{4}$
Xin-Wen\, Shu,$^{1}$
Jun-Qian\, Li,$^{1}$
\newauthor
Kun\, Jia,$^{5}$
and Han-Feng\, Song$^{6}$ \\
$^{1}$Department of Physics, Anhui Normal University, Wuhu, Anhui, 241002, China\\
$^{2}$Département d’Astronomie, Université de Genève, Chemin Pegasi 51, CH-1290 Versoix, Switzerland\\
$^{3}$Gravitational Wave Science Center (GWSC), Université de Genève, CH-1211 Geneva, Switzerland\\
$^{4}$Purple Mountain Observatory, Chinese Academy of Sciences, Nanjing 210008, China\\
$^{5}$Zhejiang University of Water Resources and Electric Power, Hangzhou, 310018, China\\
$^{6}$College of Physics, Guizhou University, Guiyang, Guizhou Province, 550025, China\\
}

\begin{document}
\label{firstpage}
\pagerange{\pageref{firstpage}--\pageref{lastpage}}
\maketitle

\begin{abstract}
IC 10 X-1, NGC 300 X-1, and Cyg X-3 constitute a unique class of X-ray binaries in which a stellar-mass black hole (BH) accretes material from a Wolf–Rayet (WR). These systems are particularly intriguing because of their short orbital periods, which make them promising progenitors of gravitational-wave (GW) sources detectable by the LIGO–Virgo–KAGRA (LVK) network. Adopting a revised accretion efficiency within the standard Bondi–Hoyle–Lyttleton framework, we perform detailed binary evolution calculations using \texttt{MESA} to characterize their properties at different evolutionary stages and to assess their ultimate fates as potential LVK-detectable GW sources. By applying additional constraints from the observed properties of IC 10 X-1 and NGC 300 X-1, we find that the upper limits on the BH masses in these systems ($M_{\rm BH} \lesssim 25\, M_\odot$ for IC 10 X-1 and $M_{\rm BH} \lesssim 15\, M_\odot$ for NGC 300 X-1) are significantly lower than previous estimates. Both systems are expected to form binary black holes (BBHs) that will merge within a Hubble time, except in the case where the BH in NGC 300 X-1 has a mass of $9\,M_\odot$, corresponding to the lower limit inferred in a previous study using the continuum-fitting method with a relativistic slim-disc model. For Cyg X-3, we find that the BH spin magnitude is constrained to be $\lesssim$ 0.6. Moreover, the WR star in Cyg X-3 is likely to form a lower-mass-gap BH, and the resulting BBH system is also expected to merge within a Hubble time.
\end{abstract}




\section{Introduction}
Wolf–Rayet (WR) stars are hot, luminous, and evolved massive stars characterized by strong emission lines of helium (He), carbon (C), nitrogen (N), and oxygen (O). These spectral features arise from their powerful, high-velocity stellar winds and significant mass-loss rates, typically on the order of $10^{-5} M_\odot$ yr$^{-1}$. WR X-ray binaries, consisting of a massive WR star paired with either a neutron star (NS) or a black hole (BH), represent a notable subclass of high-mass X-ray binaries (HMXBs). Examples include IC 10 X-1 and NGC 300 X-1 in nearby galaxies, as well as Cyg X-3 in the Milky Way (see their main properties in Table~\ref{table1}). These systems with orbital periods shorter than 1.5 d are particularly compelling, as they are immediate progenitors of double compact objects such as BBHs or BHNS systems \citep{Belczynski2012}.

Previous investigations on the evolution of IC 10 X-1/NGC 300 X-1 \citep{Bulik2011} and Cyg X-3 \citep{Belczynski2013} were carried out by using the population synthesis model \texttt{StarTrack}. A recent study utilized the open-source population synthesis tool \texttt{SEVN} to systematically investigate the role of WR-Compact Object (CO) binaries as potential progenitors of binary COs \citep{Korb2025}. Their findings suggest that approximately 70 - 100\% of Cyg X-3-like systems in a WR–BH configuration, with a BH mass $\leqslant$ 10 $M_\odot$, are likely to evolve into BCOs. Efforts have also been made to constrain the properties of IC 10 X-1, NGC 300 X-1, and Cyg X-3. Key properties of these systems, collected from the literature, are summarized in Table~\ref{table1}. These binaries are predominantly wind-fed systems, characterized by Roche lobe filling factors (defined as the ratio of a star's radius to the radius of its Roche lobe) remaining below 1.0, particularly for more massive WR stars \cite[e.g., $M_{\rm WR}$ $>$ 10 $M_\odot$,][]{Belczynski2013}. This behavior is attributed to the evolutionary shrinking of WR star radii \citep[e.g.,][]{Qin2023}. Additionally, the strong stellar winds of WR stars tend to widen the orbital separation. Consequently, no Roche lobe overflow mass transfer episodes are expected in the future evolution of these BH-WR systems \cite[see also][]{Bulik2011,Belczynski2013}. The standard Bondi-Hoyle-Lyttleton (BHL) accretion model \citep{Hoyle1939,Bondi1944,Bondi1952} is one of the most widely used frameworks for studying wind accretion in binary systems \citep{Lamers1976,Friend1982,Karino2019}. This model describes the accretion process in systems where a CO (a NS or a BH) gravitationally captures material from the stellar wind of a massive companion star. 

In this work, we utilized the detailed binary evolution code \texttt{MESA} to investigate the properties of the BH-WR binaries ( IC 10 X-1, NGC 300 X-1, and Cyg X-3) and their evolutionary fates. The paper is organized as follows. In Section~\ref{sect2}, we first introduce the main physics adopted in \texttt{MESA}. In Section~\ref{sect3}, with the properties collected from the literature, we perform detailed binary calculations to constrain their properties. Moreover, we investigate in Section~\ref{sect4} the properties of the above sources as potential GW sources. Finally, we offer a brief discussion in Section~\ref{sect5} and present our main conclusions in Section~\ref{sect6}.

\begin{table*}
    \setlength{\abovecaptionskip}{10pt} 
    \setlength{\belowcaptionskip}{10pt} 
    \centering
    \caption {Current main properties of back hole and WR X-ray binaries} 
    \renewcommand{\arraystretch}{1.4}
    \setlength{\tabcolsep}{6pt} 
    \begin{tabular}{c|c c c c c c c} 
    \hline
    \hline
        Sources & $M_{\rm BH_1}$ [$M_\odot$] & $M_{\rm WR}$ [$M_\odot$] & $P_{\rm orb}$ [d] & $\chi_1$ & $Z$ [$Z_\odot$] &$v_\infty$ $[\rm km\,s^{-1}]$ &$L_{\rm X}$ $[\rm erg\,s^{-1}]$ \\ \hline
        IC 10 X-1 & 10-30$^{(1,10)}$ & 17-35$^{(2)}$ & 1.45175$^{(3)}$ & $0.85^{(4)}$ & 0.3$^{(5,6)}$ & 1750$^{(7)}$&(1.2-4) $\times\,10^{38}$$^{(8,9)}$ \\  
        NGC 300 X-1 & 9-28$^{(10)}$ & $26_{-5}^{+7}$$^{(11)}$ & 1.3663$^{(12)}$& $0.88^{(10)} $&0.6$^{(11)}$ &1250$^{(13)}$&$1\times10^{39}$$^{(13)}$\\
        Cyg X-3 & 7.2$^{(14)}$ & $11.6_{-1.2}^{+1.2}$$^{(14)}$ & 0.2$^{(15)}$ & $/$ & 1.0$^{(16,17,18)}$ &1700$^{(19)}$&$5\times10^{38}$$^{(20)}$ \\  
        \hline
    \end{tabular}
    \begin{minipage}{\linewidth}
     \raggedright   
      \vspace{0.4cm} 
      {\small
      \textbf{References.}   (1)\cite{Wang2024}; (2)\cite{Silverman2008}; (3)\cite{Laycock2015}; (4)\cite{Steiner2016};
      (5)\cite{Massey2007};
      (6)\cite{Bulik2011};
      (7)\cite{Clark2004};
      (8)\cite{Wang2005}; 
      (9)\cite{Brandt1997}; 
      (10)\cite{Bhuvana2024};
      (11)\cite{Crowther2010};
      (12)\cite{Binder2021};    
      (13)\cite{Carpano2007A};
      (14)\cite{Antokhin2022}; (15)\cite{Parsignault1972}; (16)\cite{McCollough2016}; (17)\cite{Lemasle2018};
      (18)\cite{Korb2025}; (19)\cite{Zdziarski2012}; (20)\cite{Vilhu2009}} 
    \end{minipage} 
    \label{table1}
\end{table*}

\section{Methods}\label{sect2}
\subsection{Main physics adopted in \texttt{MESA}}
We performed the detailed binary modeling using the release version \texttt{mesa-r15140} of the Modules for Experiments in Stellar Astrophysics (\texttt{MESA}) stellar evolution code \citep{Paxton2011,Paxton2013,Paxton2015,Paxton2018,Paxton2019,Jermyn2023}. Our WR star models were constructed following the methodology outlined in recent studies \citep[i.e.,][]{Fragos2023,Hu2023,lv2023,Qin2024_gap,Qin2024_casebb}. In this work, we adopted the solar metallicity of $Z_{\odot} = $ 0.0142 as in \cite{Asplund2009}. 

We applied the mixing-length theory (MLT) \citep{MLT1958} to model convection, using a parameter value of $\alpha_{\rm mlt}=1.93$. The boundaries of the convective zone were determined using the Ledoux criterion, with a step-overshooting parameter of $\alpha_{\rm p} = 0.335\, H_{\rm p}$, calibrated based on the observed drop in rotation rates of massive main-sequence stars \citep{Brott2011}, where $H_{\rm p}$ denotes the pressure scale height at the Ledoux boundary. Semiconvection was incorporated into the WR star models following \cite{Langer1983}, with an efficiency parameter $\alpha_{\rm sc}=1.0$. We also took into account thermohaline mixing with a free parameter $\alpha_{\rm th}=1.0$ \citep{Kippenhahn1980}. For nucleosynthesis calculations, we adopted the network \texttt{approx21.net}.

We modeled rotational mixing and angular momentum transport as diffusive processes \citep{Heger2000}, accounting for the combined effects of the Goldreich–Schubert–Fricke instability, Eddington–Sweet circulations, and both secular and dynamical shear mixing. Diffusive element mixing resulting from these processes was incorporated with an efficiency parameter of $f_{\rm c}=1/30$, as proposed by \cite{Chaboyer1992,Heger2000}. To account for the sensitivity of the $\mu$-gradient to rotationally induced mixing, we mitigated its impact by reducing $f_\mu$ to a value of 0.05, following the recommendations of \cite{Heger2000}.

We adopted the mass-loss prescription outlined in \cite{Hu2022}, aligning with the recently updated models of winds from WR stars \citep{Higgins2021}. Additionally, we accounted for rotationally enhanced mass loss following the methodology described in \cite{Heger1998} and \cite{Langer1998}:
\begin{equation}\label{ml}
    \centering
    \dot{M}(\omega)= \dot{M}(0)\left(\frac{1}{1-\omega/\omega_{\rm crit}}\right)^\xi,
    \end{equation}
where $\omega$ represents the angular velocity, and $\omega_{\rm crit}$ denotes the critical angular velocity at the stellar surface. The critical angular velocity is defined as $\omega_{\rm crit}^2 = (1- L/L_{\rm Edd})GM/R^3$, where $L$, $M$, and $R$ are the star’s total luminosity, mass, and radius, respectively. The parameter $L_{\rm Edd}$ is the standard Eddington luminosity (defined using the scattering opacity of pure free electron scattering for completely ionised medium, i.e. $\kappa$ = 0.2(1 + X) $\rm cm^2$ $\rm g^{-1}$, $\rm X$ = 0 for hydrogen-free envelope of WR stars), and $G$ is the gravitational constant. The exponent $\xi = 0.43$ is adopted \cite{Langer1998}. Note that we do not include the effects of gravity-darkening, as discussed by \cite{Maeder2000}.

We applied the theory of dynamical tides to helium-rich stars with radiative envelopes, following the framework described by \cite{Zahn1977}. The synchronization timescale was calculated using the prescriptions provided in \cite{Zahn1977}, \cite{Hut1981}, and \cite{Hurley2002}, while the tidal torque coefficient $E_2$ was adopted from the updated fitting formula presented in \cite{Qin2018}. Given the previously inconsistent implementation of the synchronization timescale \citep{Sciarini2024}, we refer readers of interest to the details of the new implementation presented in \cite{Qin2024_gap}.

In our binary modeling, we evolved WR stars until their central carbon exhaustion. To calculate the baryonic remnant mass ($M_{\rm bar}$), we adopted the ``\texttt{delayed}'' SN prescription from \cite{Fryer2012}. The gravitational mass ($M_{\rm rem}$) for BHs was computed using Eq. (14) in \cite{Fryer2012}. Additionally, we accounted for the neutrino loss mechanism described by \cite{Zevin2020}. We assumed a maximum NS mass of 2.5\,$M_{\odot}$, with any object more massive being classified as a BH. We assumed direct collapse for BH formation, and as a result, the newly formed BH is not expected to experience mass loss or a natal kick \citep{Belczynski2008}.

The standard BHL accretion, however, could become inaccurate when the wind velocity ($v_{\rm w}$) is comparable to or lower than the orbital velocity ($v_{\rm orb}$), resulting in nonphysical accretion efficiencies \citep{Tejeda2025}. Thus, we adopted an updated expression \footnote{In our tests, we find that the revised accretion-efficiency prescription has a negligible impact on systems such as IC 10 X-1, NGC 300 X-1, and Cyg X-3.} for the accretion efficiency \citep[see their Eq. 16 in ][]{Tejeda2025} for circular orbits, given by:
\begin{equation}\label{eta2}
    \centering
    \eta_{\rm Tejeda24} = \left(\frac{q}{1+\omega^2}\right)^2,
\end{equation}
where $q = M2/(M_1+ M_2)$ and $w = v_w/v_{\rm orb}$. $M_1$ and $M_2$ are the donor and the accretor in the binary system, respectively.

As suggested by \cite{Bhattacharya2023}, the wind velocity at a distance $r$ from the stellar surface for a WR star can be estimated using a $\beta-$velocity law \citep{Lamers1999}, i.e.,

\begin{equation}\label{vw}
    \centering
    v_{\rm w} = v_0+(v_\infty-v_0)(1-\frac{R_1}{r})^\beta,
\end{equation}
where $v_\infty$ is the terminal velocity of WR stars, $R_1$ is the radius of the WR star, and $\beta$ is set to 1.0 for WR stars following \cite{Grafener2005,Carpano2007}. We adopted $v_\infty$ values for different systems presented in Table~\ref{table1}. Additionally, $v_0$ represents the wind velocity at the stellar surface, for which we adopt $v_0 \sim 0.01v_\infty$, as suggested for WR stars by \cite{Carpano2007}.

In close BH-WR binaries, a disk forms if the specific angular momentum of the accreted wind is sufficient to circularize outside the BH’s innermost stable circular orbit ($r_{\rm ISCO}$). Following the method of \cite{Sen2021}, we find that disks can form within the parameter space of the sources considered in this study (see Appendix A). For a disk-accreting BH, we adopt the X-ray luminosity $L_{\rm X}$ following the prescription of \cite{Wong2014},

\begin{equation}\label{lx}
    \centering
    L_{\rm X} = \eta_{\rm bol} \epsilon \frac{G M_2}{r_{\rm ISCO}} \dot{M}_{\rm acc},
\end{equation}

where $G$ is the gravitational constant, $M_2$ is the mass of the BH, and $r_{\rm ISCO}$ is the innermost stable circular orbit \citep{Bardeen1972}. The parameter $\epsilon$ represents the conversion efficiency of gravitational binding energy to radiation associated with disk accretion onto a BH \citep[with $\epsilon$ = 0.5 adopted as in][]{Belczynski2008,Wong2014}. We also adopt a bolometric correction factor $\eta_{\rm bol}$ = 0.8 for wind-fed BH systems, as suggested in \cite{Miller2001}. In this study, the BH accretes material from the stellar wind of the WR star, with mass accretion rate given by 
\begin{equation}\label{acc}
\dot{M}_{\rm acc} = \eta_{\rm Tejeda24} \dot{M}_{\rm WR},
\end{equation}
where $\dot{M_{\rm WR}}$ is mass-loss rate of the WR donor star, and $\eta_{\rm Tejeda24}$ parametrizes the efficiency of wind accretion. 

\section{Constraints on the properties of BH-WR X-ray binaries}\label{sect3}
In this section, we aim to place tighter constraints on the properties of IC 10 X-1, NGC 300 X-1, and Cyg X-3 by running separate grids of models tailored to each system. We first summarize the currently observed properties of these sources in Table~\ref{table1}, collected from the literature. The surface abundances of the WR stars in all three systems remain poorly constrained observationally. In this work, we therefore adopt a maximum relative number abundance of C/He = 0.01 for WR stars. Systems with C/He values exceeding this threshold are interpreted as being either in a more advanced evolutionary stage or overly massive, and are consequently excluded from our analysis.

\subsection{IC 10 X-1}
For the binary modeling of IC 10 X-1, we adopt initial WR star masses in the range of $17 - 45\, M_\odot$ with a step of $1.0\, M_\odot$, and initial orbital periods from 1.0 to 1.45 d, sampled uniformly in logarithmic space. We adopt $10\, M_\odot$ as the lower limit of the BH mass, motivated by recent constraints \citep{Wang2024,Bhuvana2024}.

\begin{figure*}
     \centering
     \includegraphics[width=0.48\textwidth]{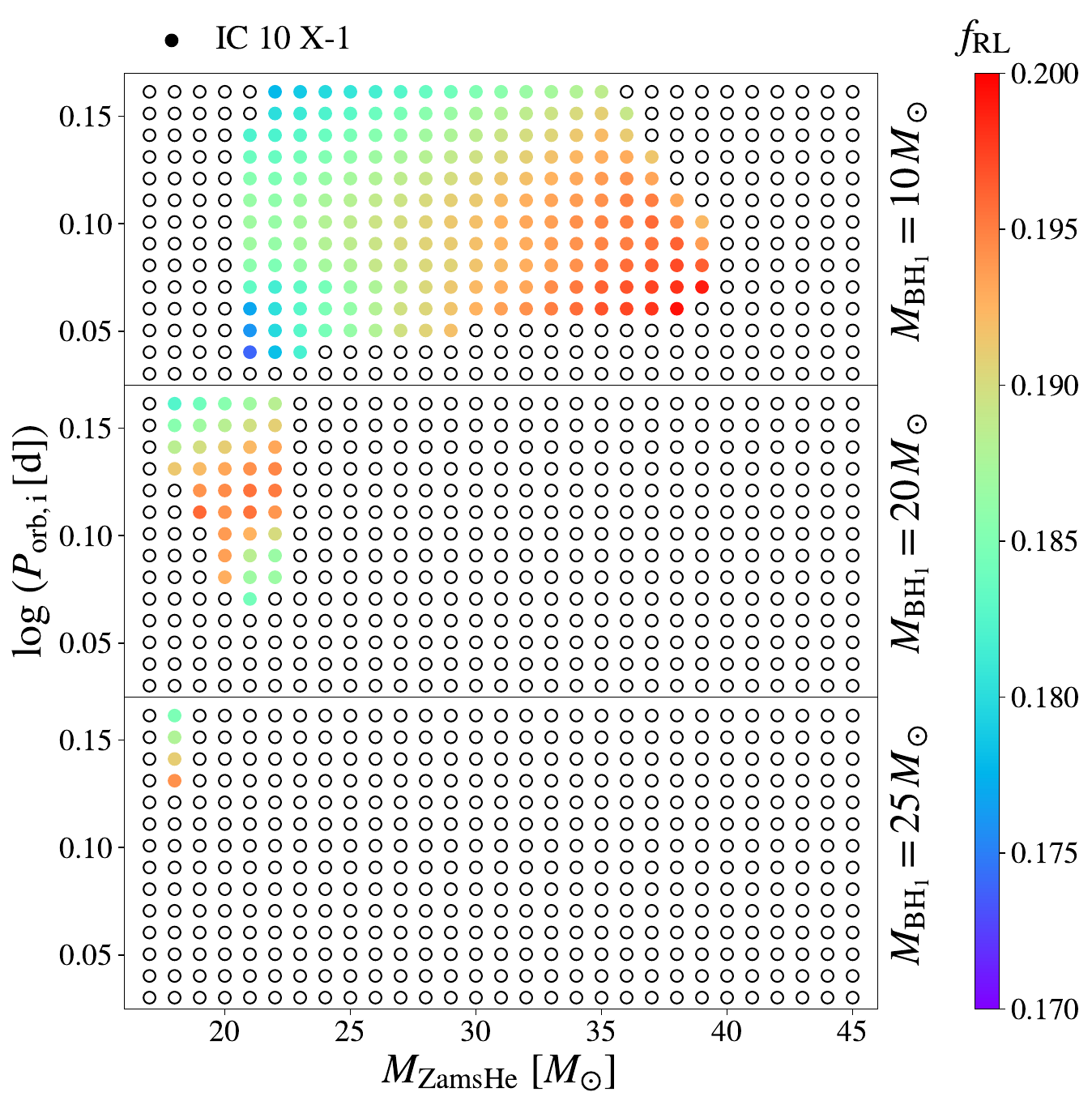}
      \includegraphics[width=0.49\textwidth]{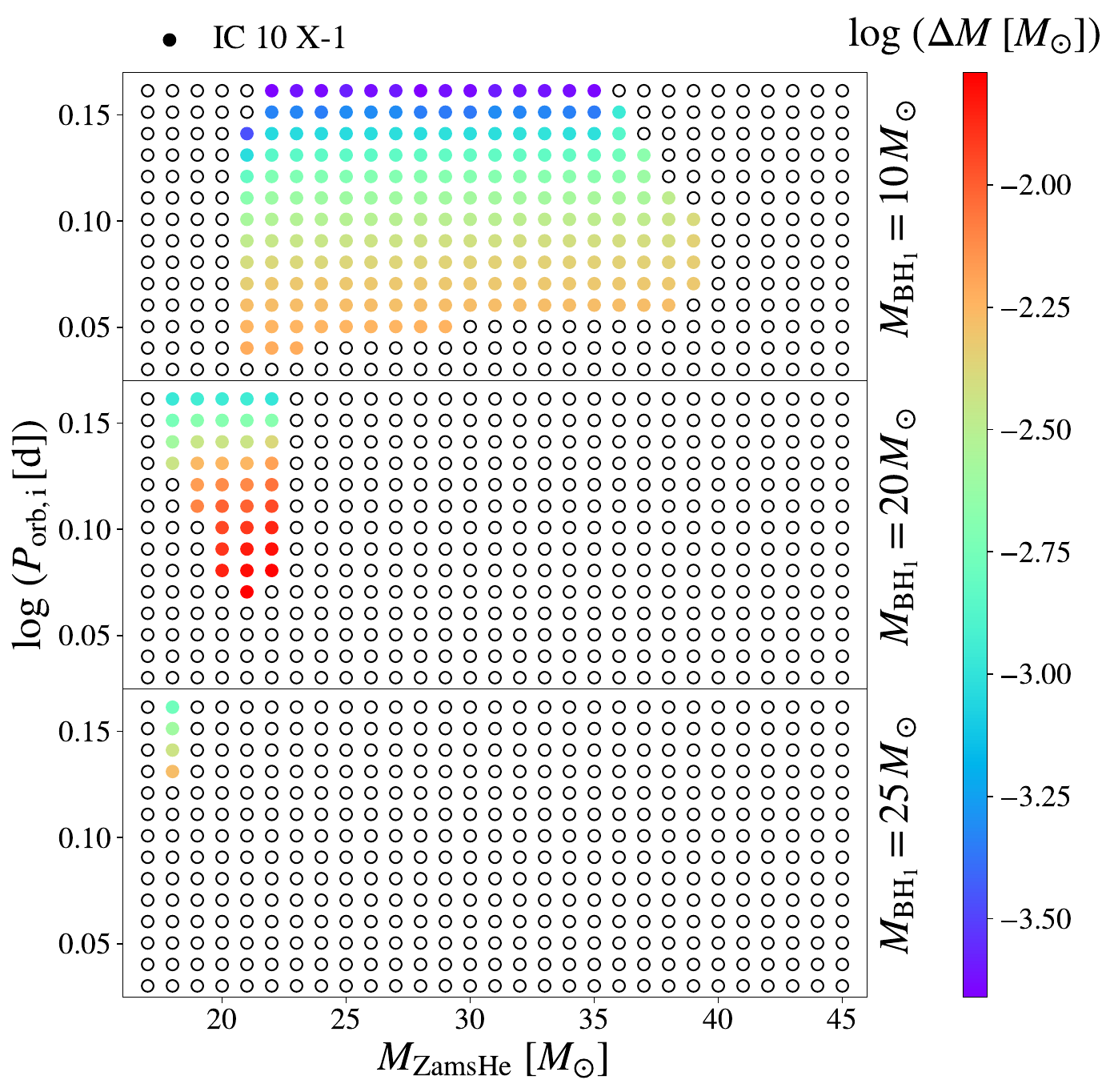}
     \caption{The Roche lobe filling factor $f_{\rm RL}$ (\textit{left panel}) and the accreted mass onto the BH (\textit{right panel}) as a function of the initial masses of the BH and WR star, as well as the initial orbital period. The colored dots indicate the parameter space likely to form the IC 10 X-1 system. Each sub-panel represents a different initial BH mass. \textit{Upper panel:} $M_{\rm BH_1} = 10\, M_\odot$, \textit{middle panel:} $M_{\rm BH_1} = 20\, M_\odot$, \textit{bottom panel:} $M_{\rm BH_1} = 25\, M_\odot$.}
     \label{ic10} 
\end{figure*}

Using a grid of detailed binary evolution, we search for the parameter space for IC 10 X-1 by matching the mass of the WR star, the orbital period, and the X-ray luminosity, $L_{\rm X}$. In the left panel of Figure~\ref{ic10} we present the Roche lobe filling factor, $f_{\rm RL}$, which quantifies how much of a star's Roche lobe is occupied by its stellar radius, as a function of different initial binary parameters. For the matched binary evolution tracks, we adopt the median value of $f_{\rm RL}$ (also for other parameters). The Roche lobe filling factor is found to be within the range of 0.17 – 0.20, with a slight dependence on the BH mass. The initial parameter space for the WR star mass shows significant sensitivity to the BH mass. Specifically, as the BH mass increases, the initial mass range of the WR star shifts to lower values, spanning from 18 $M_\odot$ to 39 $M_\odot$. Notably, the upper limit of the BH mass is constrained to approximately $25\, M_\odot$, slightly lower than earlier estimates \cite[e.g., $30\, M_\odot$ in][]{Wang2024,Bhuvana2024}. This difference is primarily attributed to current constraints on $L_{\rm X}$ (see Table~\ref{table1}). 

In the right panel, we show the mass accreted onto the BH through wind-fed accretion. The accreted mass ranges from $\sim 2.2 \times 10^{-4} M_\odot$ to $\sim 6.1 \times 10^{-3} M_\odot$ for a $10\, M_\odot$ BH accretor (from $\sim 1.0 \times 10^{-3} M_\odot$ to $\sim 1.7 \times 10^{-2} M_\odot$ for a $20\, M_\odot$ BH accretor, and from $\sim 1.7 \times 10^{-3} M_\odot$ to $\sim 5.3 \times 10^{-3} M_\odot$ for a $25\, M_\odot$ BH accretor). These values depend on the initial orbital period and WR star mass. Given the relatively small accreted mass, the BH spin is not significantly increased \citep[see recent findings in][]{Qin2024a, Qin2024d}. The current mass range of the WR stars is constrained to be in a range of $\sim 17$ -- $\sim 35\, M_\odot$ (depending on the BH mass), consistent with previous estimates by \cite{Silverman2008}. 

\subsection{NGC 300 X-1}
To model NGC 300 X-1, we cover an initial WR star mass range of $21 -45\, M_\odot$ with increments of $1.0\, M_\odot$ and an initial orbital period spanning from approximately $\sim 0.9$ d to 1.36 d, distributed logarithmically. Based on recent constraints from \cite{Bhuvana2024}, we adopt $9\, M_\odot$ as the lower limit for the BH mass. Given the proportional relationship between X-ray luminosity and BH mass (see Eq.~\ref{lx}), we iteratively decrease the BH mass until a model from our grid aligns with the observed system properties. The upper limit for the BH mass is determined to be around $15\, M_\odot$.

\begin{figure*}
     \centering
     \includegraphics[width=0.48\textwidth]{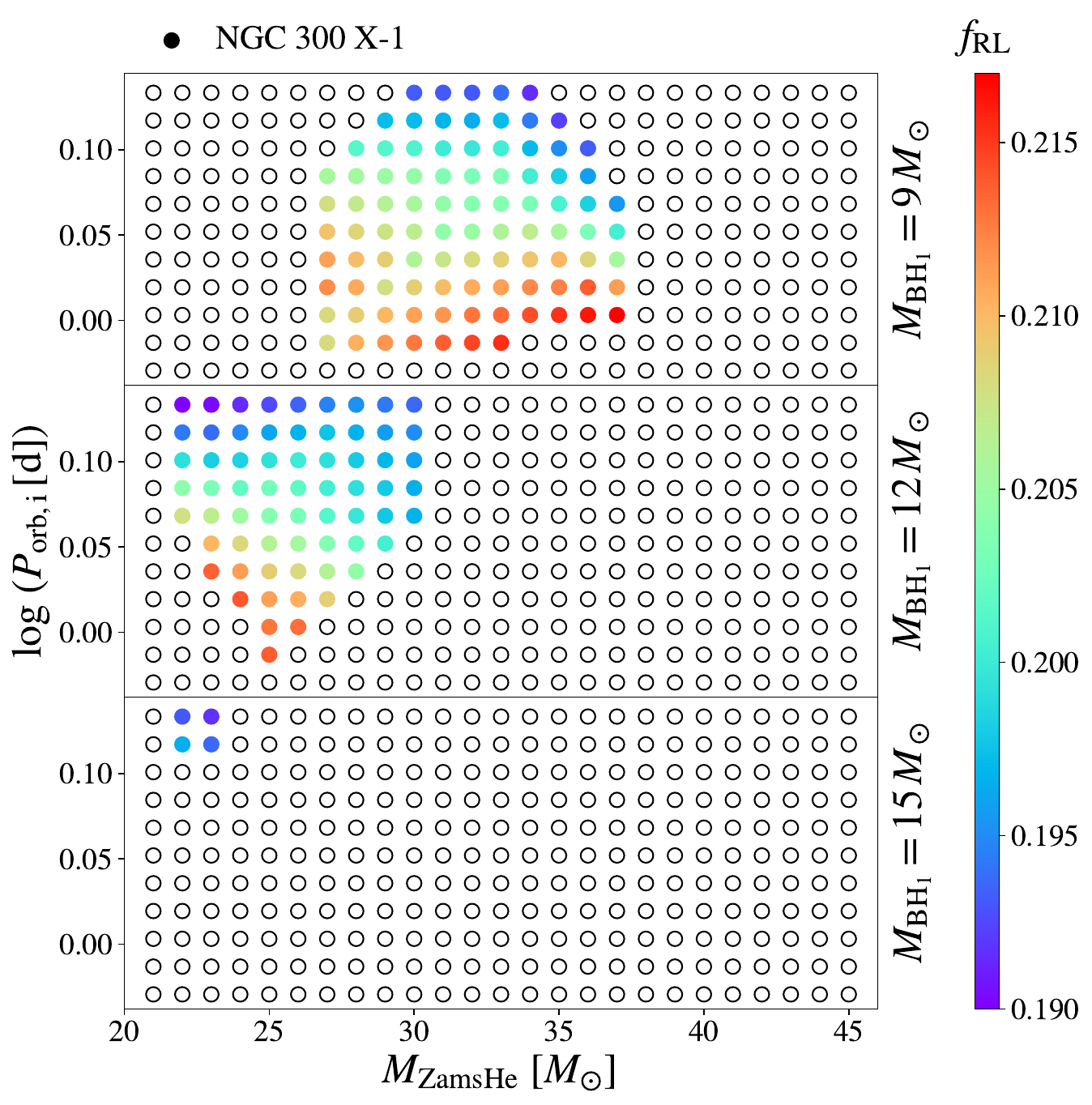}
     \includegraphics[width=0.49\textwidth]{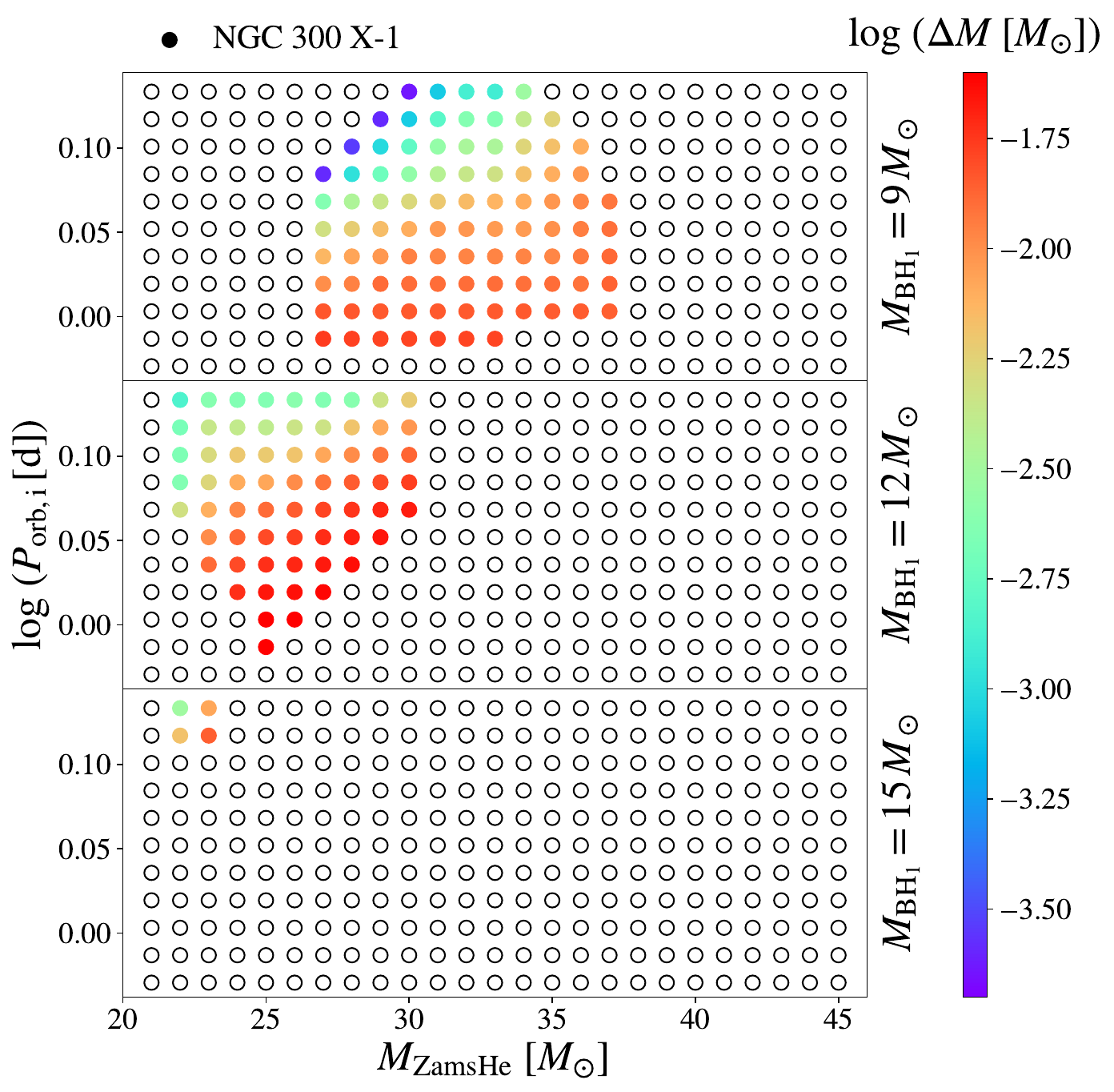}
     \caption{Similar to Figure~\ref{ic10}, but for NGC 300 X-1.}
     \label{ngc300} 
\end{figure*}

In the left panel of Figure~\ref{ngc300}, we present the Roche lobe filling factor ($f_{\rm RL}$) as a function of the initial conditions likely leading to the formation of NGC 300 X-1. The $f_{\rm RL}$ values range between 0.19 and 0.22, slightly larger than those for IC 10 X-1. Similar to IC 10 X-1, the lower limit of the initial orbital period is approximately 1.0 d. Moreover, the upper limit of the initial WR star mass decreases with increasing BH mass, shifting to 37 $M_\odot$ for $M_{\rm BH_1} = 9\, M_\odot$ (12 $M_\odot$ for $M_{\rm BH_1} = 30\, M_\odot$ and 23 $M_\odot$ for $M_{\rm BH_1} = 15\, M_\odot$). Similar to IC 10 X-1, the parameter space for forming NGC 300 X-1 shrinks as the BH companion mass increases. The right panel shows the mass accreted onto the BH, ranging from $\sim 2.3\, \times 10^{-4} M_\odot$ to $\sim 1.7\, \times 10^{-2} M_\odot$ for a $9\, M_\odot$ BH accretor (from $\sim 1.4\, \times 10^{-3} M_\odot$ to $\sim 2.8\, \times 10^{-2} M_\odot$ for a $12\, M_\odot$ BH accretor and from $\sim 3.0\, \times 10^{-3} M_\odot$ to $\sim 1.4\, \times 10^{-2} M_\odot$  for a $15\, M_\odot$ BH accretor).

\subsection{Cyg X-3}
Cyg X-3 is currently the only confirmed X-ray binary system in our galaxy comprising a WR star and a compact object, but the exact nature of the compact object remains under debate. Recent constraints from \cite{Nie2024} favor the compact object being a BH rather than a NS. Additionally, the mass of the BH has been tightly constrained to approximately $7.2\, M_\odot$ \citep{Antokhin2022}, a value that will be used in the subsequent modeling. Compared to IC 10 X-1 and NGC 300 X-1, Cyg X-3 features more precise constraints on the masses of its two components and a notably shorter orbital period (see Table~\ref{table1}). 

In our binary modeling, we adopt the initial WR star mass in a range of $10 -20\, M_\odot$ with a step of $1.0\, M_\odot$, and initial orbital periods spanning from $\sim 0.1$ d to $\sim 0.2$ d, in a logarithmic space. Notably, the observed X-ray luminosity ($L_x = 5 \times 10^{38} \rm erg\,s^{-1}$) provides a critical constraint on the spin magnitude of the BH. To determine this, we vary the initial spin of the BH starting from 0.1, adjusting until the X-ray luminosity matches the observed value using Eq.~\ref{lx}. We find that the BH spin magnitude cannot exceed 0.6. To further refine our model, we conduct separate grids assuming BH spin magnitudes of 0.1 and 0.6, respectively, allowing us to better constrain the other properties of Cyg X-3.

\begin{figure*}
     \centering
     \includegraphics[width=0.47\textwidth]{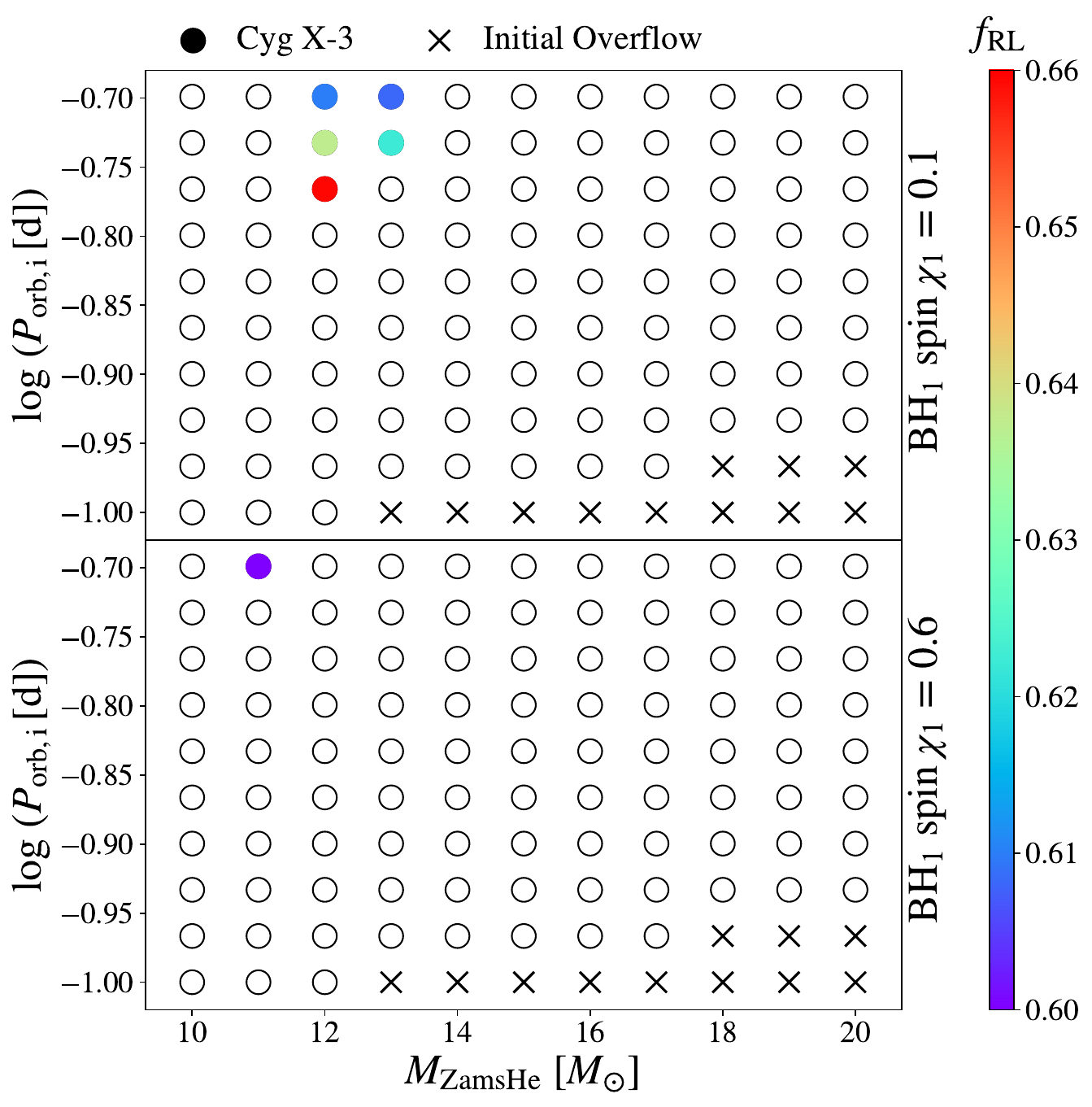}
     \includegraphics[width=0.49\textwidth]{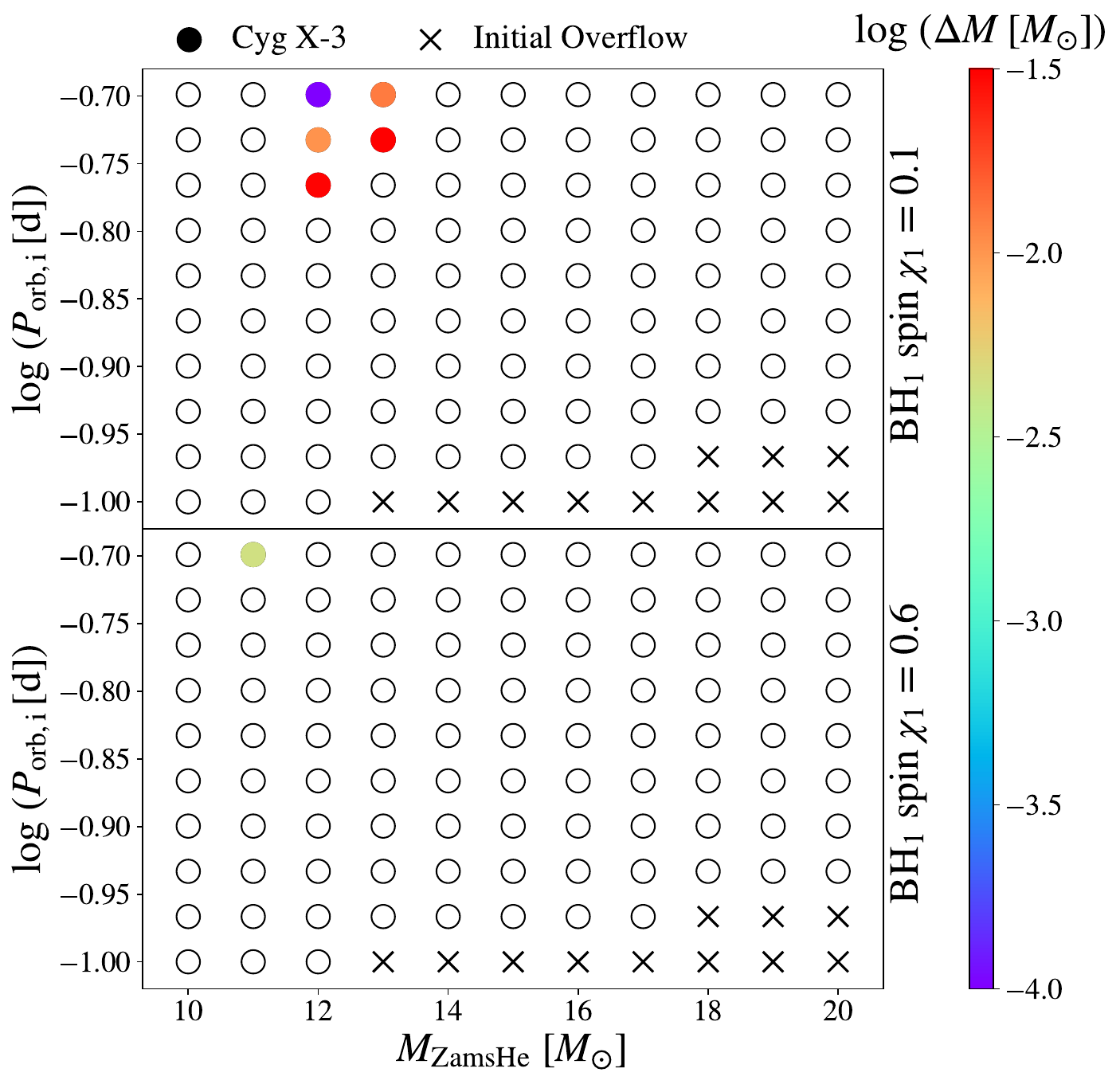}
     \caption{Similar to Figure~\ref{ic10}, but for Cyg X-3.}
     \label{cygx3} 
\end{figure*}

In the left panel of Figure~\ref{cygx3}, we show that the Roche lobe filling factor ($f_{\rm RL}$) for Cyg X-3 ranges from 0.60 to approximately 0.66, which is significantly higher than the values found for IC 10 X-1 and NGC 300 X-1. The initial mass of the WR star is constrained to a narrow range of 11 - 12 $M_\odot$, and the corresponding orbital period has a lower limit of approximately 0.17 d (see the right panel). The accreted mass varies from $\sim\,8.6\, \times 10^{-5} M_\odot$ to $\sim 3.2\, \times 10^{-2} M_\odot$ for the BH spin $\chi_1 = 0.1$ ($\sim 4.4\, \times 10^{-3} M_\odot$ for the BH spin $\chi_1 = 0.6$). 

Since the mass accreted onto the BH is limited (see the right panel), the current mass of the BH likely reflects its mass at birth. For IC 10 X-1 and NGC 300 X-1, we find that the current mass of the WR star is consistent with earlier findings (as shown in Table~\ref{table1}). For Cyg X-3, we determine that the current WR star has a mass in the range of 10.6 - 12.3 $M_\odot$ when the BH spin is $\chi_1 = 0.1$, with a slightly higher mass of approximately 10.8 $M_\odot$ for a BH spin of $\chi_1 = 0.6$. 

\subsection{Best-fit models}
\begin{figure}
     \centering
     \includegraphics[width=0.45\textwidth]{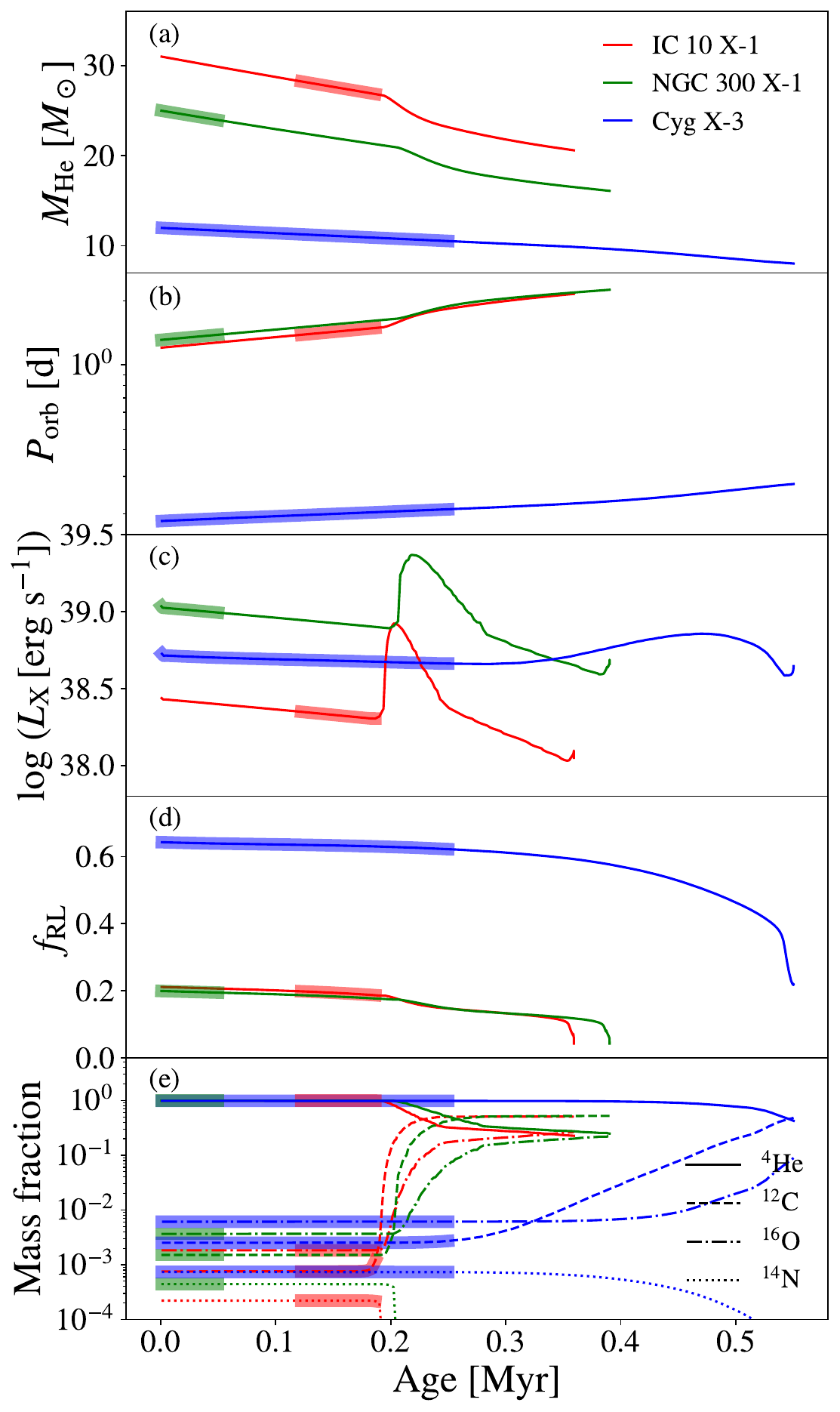}
     \caption{Evolutionary tracks of the best-fit models for IC 10 X-1 (red), NGC 300 X-1 (green), and Cyg X-3 (blue). The panels display the WR star mass $M_{\rm He}$ (a), orbital period $P_{\rm orb}$ (b), X-ray luminosity $L_{\rm X}$ (c), Roche-lobe overflow factor $f_{\rm RL}$ (d), and surface abundances (e). In panel (e), solid, dashed, dash-dotted, and dotted lines denote $^4$He, $^{12}$C, $^{16}$O, and $^{14}$N, respectively. The bold shaded regions indicate the current plausible evolutionary states of each system.}
     \label{best_fit} 
\end{figure}

In this section, we present the best-fit binary evolutionary sequences for IC 10 X-1, NGC 300 X-1, and Cyg X-3. These sequences are identified by searching our model grid for tracks that best reproduce the observed properties of each system, namely the He-star mass ($M_{\rm He}$), orbital period ($P_{\rm orb}$), and X-ray luminosity ($L_{\rm X}$) (see Table~\ref{table1}). Figure~\ref{best_fit} displays the corresponding evolutionary tracks, including the past, present, and future stages. The present properties of each best-fit model are highlighted in shading.

For IC 10 X-1, we find $M_{\rm He} = 26.8$ -- $28.3\,M_\odot$, $P_{\rm orb} = 1.39$ --$1.50\,{\rm d}$, $L_{\rm X} = (2.08$ -- $2.24)\times10^{38}\,{\rm erg s^{-1}}$, and $f_{\rm RL}$ = 0.19 -- 0.20.
For NGC 300 X-1, the corresponding values are $M_{\rm He} = 23.95$ --$25.0\,M_\odot$, $P_{\rm orb} = 1.31$ -- $1.39\,{\rm d}$, $L_{\rm X} = (9.84\times10^{38}$ -- $1.08\times10^{39})\,{\rm erg s^{-1}}$, and $f_{\rm RL}$ = 0.19 -- 0.20. For Cyg X-3, we obtain $M_{\rm He} = 10.55$ -- $12.0\,M_\odot$, $P_{\rm orb}$ = 0.19 -- 0.21$\,{\rm d}$, $L_{\rm X}$ = (4.70 -- 5.31)$\times10^{38}\,{\rm erg s^{-1}}$, and $f_{\rm RL}$ = 0.60 -- 0.63.

The surface chemical abundances of the best-fit models are also shown. For IC 10 X-1, we find $^4$He = $(9.94$ -- $9.95)\times10^{-1}$, $^{12}$C = $(7.55\times10^{-4}$–$2.48\times10^{-3})$, $^{16}$O = $(1.83$ -- $1.84)\times10^{-3}$, and $^{14}$N = $(2.01$ -- $2.21)\times10^{-4}$.
For NGC 300 X-1, we obtain $^4$He $\simeq 9.91\times10^{-1}$, $^{12}$C $\simeq 1.51\times10^{-3}$, $^{16}$O $\simeq 3.66\times10^{-3}$, and $^{14}$N $\simeq 4.42\times10^{-4}$.
For Cyg X-3, we find $^4$He $\simeq 9.85\times10^{-1}$, $^{12}$C = $(2.51$ -- $2.72)\times10^{-3}$, $^{16}$O $\simeq 6.10\times10^{-3}$, and $^{14}$N = $(7.34$ -- $7.37)\times10^{-4}$.

\section{Properties as potential gravitational-wave sources}\label{sect4}
IC 10 X-1, NGC 300 X-1, and Cyg X-3 are likely immediate progenitors of binary black holes, representing outcomes of classical isolated massive binary evolution. With the initial properties of BH-WR systems at formation now well-constrained, we can advance to predicting their potential evolution and final outcomes using detailed binary modeling.

As previously mentioned, we adopt the ``\texttt{delayed}'' SN prescription \citep{Fryer2012} to calculate the gravitational mass of the BH. For simplicity, we assume that WR stars evolving up to carbon depletion will soon directly collapse to form BHs without any significant loss of angular momentum. It is noteworthy that the ``\texttt{delayed}'' mechanism can produce BHs with masses in the lower mass gap \cite[e.g., 2.5 -- 5 M$_\odot$, see][]{Fryer2012}. We present the mass and spin magnitude of the resulting BH for the three sources in Table~\ref{table2}. The chirp mass, $M_{\rm chirp}$, is defined as:
    \begin{equation}\label{chirp}
    \centering
    M_{\rm chirp} = \frac{(M_{\rm 1} M_2)^{3/5}}{(M_{\rm 1}+M_2)^{1/5}},
    \end{equation}
where $M_1$ and $M_2$ are the masses of the two BHs, respectively. The effective inspiral spin ($\chi_{\rm eff}$), which can be directly constrained by the GW signal, is defined as:
\begin{equation}\label{eq1}
    \chi_{\rm eff}=\frac{M_1\chi_{1} + M_2\chi_{2}}{M_1+M_2},
\end{equation}
where $\chi_{1}$ and $\chi_{2}$ are the corresponding dimensionless spin parameters of the two BHs, aligned with the orbital angular momentum. Following the formation of double BHs, the emission of gravitational waves gradually shrinks their orbits by removing orbital angular momentum, ultimately driving the merger of the two BHs. To calculate the merger time for these binary BHs, we use the expression provided by \cite{Peters1964}:
\begin{equation}
    T_{\rm merger}=\frac{5}{256} \frac{c^{5} a_{\rm f}^{4}}{G^{3} (M_1 + M_2)^{2} M_{\rm r}} T(e) , 
\end{equation}
where $c$ is the speed of light and $M_r$ is the binary’s reduced mass, $M_1$ and $M_2$ are the first-born and second-born BHs, respectively, and $a_{\rm f}$ is the orbital separation between the two components. For simplicity, we assume that the BHs form in circular orbits, i.e., $T(e) = 1$.

\begin{figure}
     \centering
     \includegraphics[width=0.49\textwidth]{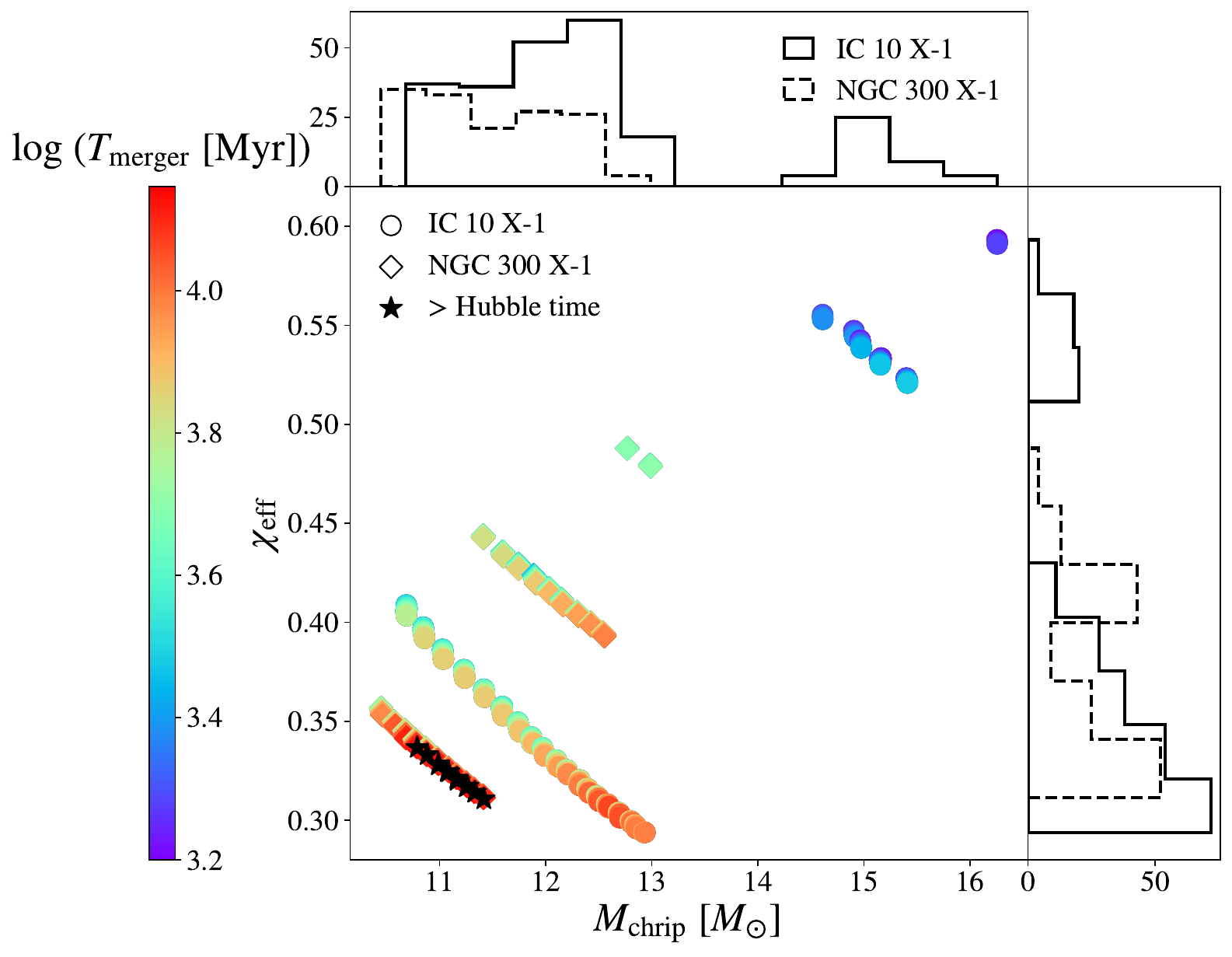}
     \caption{$T_{\rm merger}$ as a function of $\chi_{\rm eff}$ and $M_{\rm chirp}$ for IC 10 X-1 and NGC 300 X-1, respectively. \textit{Upper panel:} distribution of $M_{\rm chirp}$, \textit{right panel:} distribution of $\chi_{\rm eff}$. Circles: IC 10 X-1, diamonds: NGC 300 X-1. Black stars represent the systems that will not merge within the Hubble time.}
     \label{Tmerger_IC-NGC} 
\end{figure}

In Figure~\ref{Tmerger_IC-NGC}, we first present $T_{\rm merger}$ for IC 10 X-1 (circles) as a function of $M_{\rm chirp}$ and $\chi_{\rm eff}$. $T_{\rm merger}$ is found to range from approximately $\sim 10^{9.2}$ yr to $\sim 10^{10.1}$ yr (log($T_{\rm merger}$[Myr]) $\in [\sim 3.20, \sim 4.06]$), which is shorter than Hubble time. We can see that $M_{\rm chirp}$ and $\chi_{\rm eff}$ are highly sensitive to the assumed mass of the BH companion. Specifically, $M_{\rm chirp}$ varies from $\sim 10.7\, M_\odot$ to $\sim 16.3\, M_\odot$ and $\chi_{\rm eff}$ spans from $\sim 0.3$ to $\sim 0.6$. A similar trend is obeserved for NGC 300 X-1 (diamonds in Figure~\ref{Tmerger_IC-NGC}), where $T_{\rm merger}$ ranges from $\sim 10^{9.5}$ yr to $\sim 10^{10.3}$ yr (log($T_{\rm merger}$[Myr]) $\in [\sim 3.48, \sim 4.26]$). Notably, some systems with a BH companion mass of 9 $M_\odot$ are unable to merge within the Hubble time.

\begin{figure*}
  \centering
  \begin{minipage}[b]{1\textwidth}
    \centering  
    \includegraphics[width=0.9\textwidth]{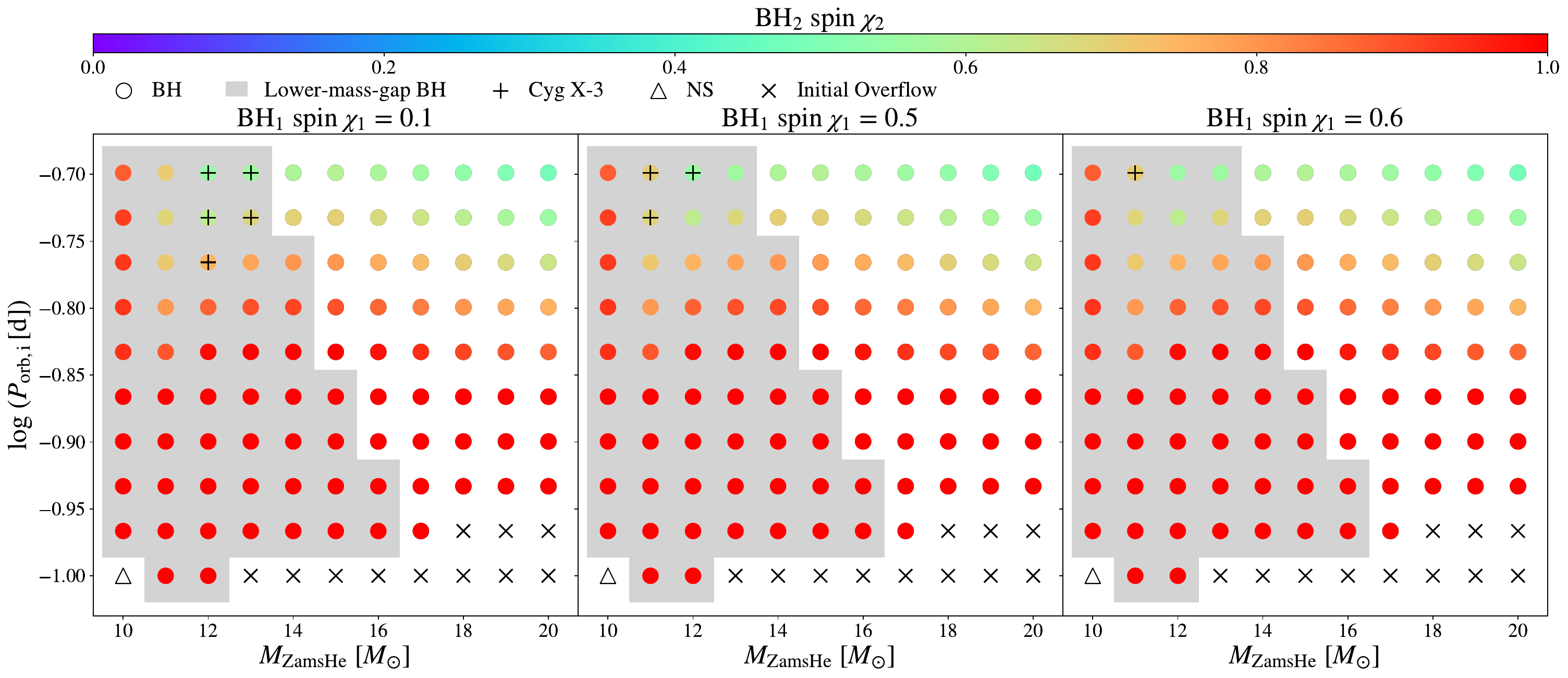}
    \caption{The spin (the color bar) of the second-born BH ($BH_{\rm 2}$) formed through direct core-collapse of WR star, as a function of the initial WR star mass and the initial orbital period. \textit{Left panel:} the first-born $BH_1$ spin $\chi_1 = 0.1$, \textit{middle panel:} the first-born $BH_1$ spin $\chi_1 = 0.5$, \textit{right panel:} the first-born $BH_1$ spin $\chi_1 = 0.6$. Circles: BH, plus: the systems that likely resemble Cyg X-3, triangles: NS, crosses: the systems that initially overflow their Roche lobes. The shaded regions represent the parameter space where the WR star is likely to form a BH that falls within the so-called lower-mass gap.}
    \label{cygx3-spin}
  \end{minipage}
\end{figure*}

Interestingly, for Cyg X-3, we find that the WR star is more likely to form a mass-gap BH with a mass between 3.9 $M_\odot$ and 4.6 $M_\odot$. This system resides in a very short orbit, where the tidal effects on the WR star are significantly strong, leading to the formation of a fast-spinning BH with $\chi_2 \in [\sim0.55, \sim0.75]$ (see Figure~\ref{cygx3-spin}, we also include an additional grid assuming a dimensionless spin of 0.5 for the first-born BH). The chirp mass $M_{\rm chirp}$ and effective inspiral spin $\chi_{\rm eff}$ for this system are found to range from $\sim 4.6\, M_\odot$ to $\sim 5.0\, M_\odot$ and from $\sim 0.3$ to $\sim 0.6$, respectively. Additionally, this system is likely to form a potential BBH with a shorter merging timescale of log($T_{\rm merger}$[Myr]) $\in [\sim 2.17, \sim 2.31]$. Notably, it is worth emphasizing that the spin magnitude of the first-born BH in this system should not exceed 0.6. In Table~\ref{table2}, we briefly summarize the main properties of BH and WR X-ray binaries at various evolutionary phases.

\begin{table*}
    \setlength{\abovecaptionskip}{10pt} 
    \setlength{\belowcaptionskip}{10pt} 
    \centering
    \caption{Updated main properties of BH-WR X-ray binaries} 
    \renewcommand{\arraystretch}{1.4} 
    \resizebox{\textwidth}{!}{%
    \begin{tabular}{c|c c | c c c| c c|c c| c c c} 
    \hline\hline
    \multirow{2}{*}{Sources} & \multicolumn{2}{c|}{He-Zams} & \multicolumn{3}{c|}{Present stage} &\multicolumn{2}{c|}{He-Tams} & \multicolumn{2}{c|}{Formation of BBHs} & \multicolumn{3}{c}{Properties as GW sources} \\ 
    \cline{2-13}
    & $M_{\rm ZamsHe}$ [$M_\odot$] & $P_{\rm orb,i}$ [d] & $f_{\rm RL}$ & $M_{\rm BH_1}$ [$M_\odot$] & $\chi_1$ & $M_{\rm TamsHe}$ [$M_\odot$]& $P_{\rm orb}$ [d] & $M_{\rm BH_2}$ [$M_\odot$] & $\chi_2$ & $M_{\rm chrip}$ [$M_\odot$] & $\chi_{\rm eff}$ & log ($T_{\rm merger}$ [Myr]) \\ \hline
    IC 10 X-1 & 18-39 & 1.10-1.45 & 0.17-0.20 & $\leq$ 25 & / & 14.8-23.5&1.50-2.82&14.2-22.8 & 0.05-0.14 & 10.7-16.3 & 0.3-0.6 & 3.20-4.06\\  
    NGC 300 X-1 & 22-37 & 0.93-1.36 & 0.19-0.22 & $\leq$ 15 & / & 14.9-20.3 &1.65-3.10&14.3-19.6 & 0.05-0.07 & 10.4-13.0 & 0.3-0.5 & 3.48-4.26\\
    Cyg X-3 & 11-12 & 0.17-0.20 & 0.60-0.66 & / & $\leq$ 0.6 &8.1-8.7&0.25-0.30& 3.9-4.6 & 0.55-0.75 & 4.6-5.0 & 0.3-0.6 & 2.17-2.31\\  
    \hline
    \end{tabular}%
    }
    \begin{minipage}{\linewidth}  
      \vspace{0.4cm} 
      \justifying  
      {\small The slash symbol indicates properties on which no constraints are imposed in this work. Helium Zero-Age Main Sequence (He-Zams): onset of core helium burning; Helium Terminal-Age Main Sequence (He-Tams): end of core helium burning.}
    \end{minipage} 
    \label{table2}
\end{table*}

\section{Discussion}\label{sect5}

It is important to note that our constraints on the three sources are subject to some uncertainties. In particular, X-ray luminosity plays a crucial role in setting the upper limit on the current BH mass for IC 10 X-1 and NGC 300 X-1, as well as the BH spin magnitude for Cyg X-3. Earlier investigations of O-star donors suggested that, despite high accretion rates, the angular momentum of the accreted wind is generally insufficient to form an accretion disk around the BH \citep{Illarionov1975,Hirai2021,Sen2021,Sen2024}. However, X-ray luminosities exceeding $\rm 10^{37} erg\, s^{-1}$ are typically associated with BH-WR systems, where an accretion disk can form around the BH \citep{Sen2025}. For instance, the X-ray spectrum of IC 10 X-1 is strongly disk-dominated \citep{Steiner2016}. Recent observations of NGC 300 X-1 are consistent with the presence of a wind–Roche lobe overflow accretion disk. 

In another class of wind-fed HMXBs (Cygnus X-1, LMC X-1, and M33 X-7), the BHs are observed to have extremely high spins, with the donor star being of OB type. The above wind-fed HMXBs are likely immediate progenitors of BH–WR systems. A recent study from \citep{Xing2025} explored the formation of these systems using the binary population synthesis framework \texttt{POSYDON} \citep{Fragos2023}, which builds on \texttt{MESA}-based stellar grids at solar metallicity. They concluded that reproducing the observed high BH spins requires assuming highly conservative accretion (i.e., most or all of the mass lost by the donor star is accreted by the companion, rather than being lost from the system), regardless of the specific accretion prescription.

Additionally, the accretion rate is directly influenced by the wind mass-loss rates of WR stars, making it a key source of uncertainty. To account for this, we adopt the standard ``\texttt{Dutch}'' scheme wind prescription \citep{Nugis2000}, and apply a scaling factor of 2/3 to align with the WR wind models \citep{Higgins2021}. However, we recognize that this prescription does not fully account for the complexities of WR winds \citep{Sander2020}, including clumping, metallicity dependence, and anisotropies, etc. We did not explore the impact of changing the prescription for the WR winds.

\section{Conclusions}\label{sect6}

In this work, we perform binary evolution calculations to constrain the formation parameter space of BH–WR systems, using the observed properties of IC 10 X-1, NGC 300 X-1, and Cyg X-3 collected from the literature. This allows us to systematically investigate their evolutionary pathways and assess their potential as gravitational-wave sources detectable by the LVK network.

Using a revised accretion-efficiency prescription, we first find that it has a negligible impact on IC 10 X-1, NGC 300 X-1, and Cyg X-3. With constraints from the observation, we further find that the BH companion mass of the WR star in IC 10 X-1 should not exceed $25\, M_\odot$. Notably, the mass of the BH plays a crucial role in determining the mass range of the WR stars. The accreted mass onto the BH is found to range from approximately 2 $\times 10^{-4} M_\odot$ to 1.7 $\times 10^{-2} M_\odot$, depending on the BH mass. For NGC 300 X-1, the upper limit of the BH mass is constrained to be around $15\, M_\odot$, which is significantly lower than the earlier estimates (See Table~\ref{table1}). The accreted mass onto the BH is less than $0.02\, M_\odot$ (comparable to IC 10 X-1), which is insufficient to significantly spin up the BH. As the only known BH-WR system in our galaxy, the BH mass in Cyg X-3 is well-constrained, but its spin magnitude remains uncertain. Our findings suggest that the spin magnitude of the BH ($\chi_1$) is unlikely to exceed $\sim 0.6$, lower than those reported for IC 10 X-1 \citep{Steiner2016} and NGC 300 X-1 \citep{Bhuvana2024}.

Table~\ref{table2} summarizes the key properties of BH and WR X-ray binaries constrained by this study across different evolutionary phases. As the initial parameter space of the BH-WR systems for IC 10 X-1, NGC 300 X-1, and Cyg X-3 has been constrained, we further explore their properties as GW sources. First, both IC 10 X-1 (log($T_{\rm merger}$[Myr]) $\in [\sim 3.20, \sim 4.06]$) and Cyg X-3 (log($T_{\rm merger}$[Myr]) $\in [\sim 2.17, \sim 2.31]$) are expected to evolve into BBHs that will merge within the Hubble time. NGC 300 X-1 is expected to form a BBH; however, if the mass of its first-born BH is assumed to be $9\, M_\odot$, the system will not merge within the Hubble time. Interestingly, the WR star in Cyg X-3 is predicted to form a BH that falls within the so-called lower-mass gap. Consequently, this system may evolve into a BHNS binary detectable by the LVK network.

For IC 10 X-1, the chirp mass is expected to range from approximately 10.7 $M_\odot$ to 16.3 $M_\odot$ (NGC 300 X-1: $\sim 10.4\, - \sim 13.0\, M_\odot$, Cyg X-3: $\sim 4.6\, - \sim 5.0\, M_\odot$). Additionally, the effective inspiral spin $\chi_{\rm eff}$ for IC 10 X-1 is found to be in the range of $\sim 0.3\, - \sim\, 0.6$ (NGC 300 X-1: 0.3 - 0.5, Cyg X-3: 0.3 - 0.6).

\section*{Acknowledgements}
We thank the anonymous referee for helpful comments that improved this manuscript. This is work is supported by the National Natural Science Foundation of China (grant Nos. 12473036 and 12573045) and Anhui Provincial Natural Science Foundation (grant No. 2308085MA29). G.M. has received funding from the European Research Council (ERC) under the European Union’s Horizon 2020 research and innovation program (grant agreement No 833925, project STAREX). Q.Z. Liu is supported by the National Key R\&D program of China (2021YFA0718500), and the National Natural Science Foundation of China under Grants No. 12473042 and 12233002. This work was supported by Anhui Province Graduate Education Quality Engineering Project (grant No. 2024qyw/sysfkc012) and by Jiangxi Provincial Natural Science Foundation (grant Nos. 20242BAB26012 and 20224ACB211001). H.F. Song is supported by the National Natural Science Foundation of China (grant Nos. 12173010 and 12573034). All figures are made with the free Python module Matplotlib \citep{Hunter2007}.

\section*{Data availability}
The data generated in this work will be shared
upon reasonable request to the corresponding author.


\bibliographystyle{mnras}
\bibliography{ref} 
\appendix
\section{Disk formation}
To determine whether an accretion disk can form around the BH, we adopt the methodology outlined in \cite{Sen2021}, as described below:
\begin{equation}\label{}
    \centering
    \frac{R_{\rm disk}}{R_{\rm ISCO}} > 1,
\end{equation}
where $R_{\rm disk}$ and $R_{\rm \rm ISCO}$ are the circularization radius of the accretion disk and the radius of the innermost stable orbit, respectively. The disk formation criterion can be further expressed as:
\begin{equation}\label{eq_criterion}
    \centering
    \frac{R_{\rm disk}}{R_{\rm ISCO}}=\frac{2}{3}\frac{\eta_{\rm nf}^2}{(1+q)^2}\left(\frac{v_{\rm orb}}{c}\right)^{-2}\left(1+\frac{v^2_w}{v^2_{\rm orb}}\right)^{-4}\gamma^{-1}_{\rm \pm}>1,
\end{equation}
where $\eta_{\rm nf}$ denotes the efficiency of specific angular momentum accretion by the BH, $q$ represents the mass ratio ($q = M_{\rm WR}/M_{\rm BH}$), and $\gamma_{\rm \pm}$ accounts for the modification of the innermost stable circular orbit radius induced by the BH spin. We adopt the efficiency factor $\eta_{\rm nf} = 1/3$ derived from detailed hydrodynamical simulations \citep{Livio1986,Ruffert1999}. As $\gamma_{\rm \pm}$ varies from 1/6 for a maximally spinning BH with a prograde disk to 3/2 for a maximally spinning BH with a retrograde disk, assuming alignment between the disk and the BH's angular momentum \citep{Ileyk2017}. In this work, we adopt the lower limit of $\gamma_{\rm \pm} = 1/6$, which corresponds to the minimum values of $R_{\rm disk}/R_{\rm \rm ISCO}$.

\begin{figure}
     \centering
     \includegraphics[width=0.435\textwidth]{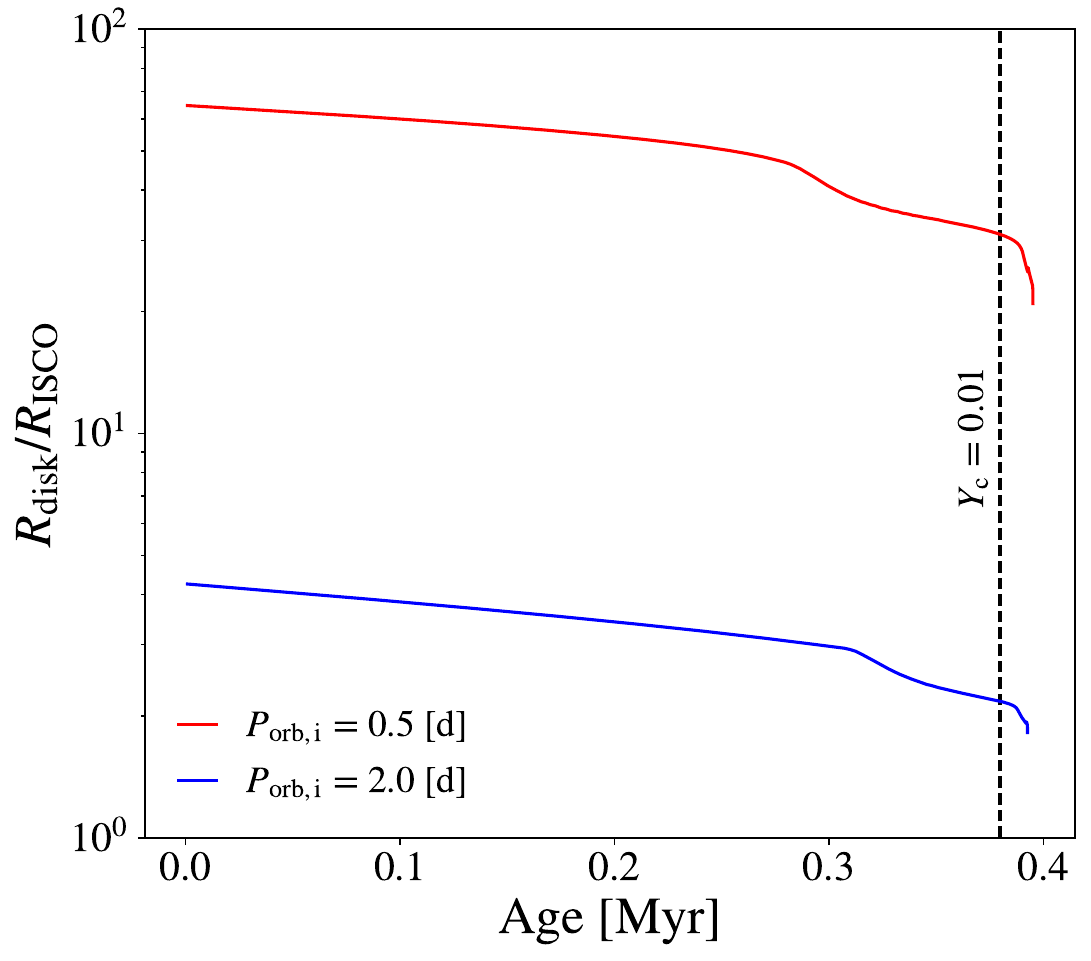}
     \caption{The ratio $R_{\rm disk}/R_{\rm \rm ISCO}$ as a function of the WR star's age for different initial orbital periods (red solid line: $P_{\rm orb,i} = 0.5$ d, blue solid line: $P_{\rm orb,i} = 2.0$ d). The binary system evolves until central carbon depletion, assuming initial masses of 23 $M_\odot$ for the WR star and 20 $M_\odot$ for the BH. The vertical dashed line indicates the point of central helium depletion.}
     \label{Rdisk1} 
\end{figure}

\begin{figure}
     \centering
     \includegraphics[width=0.45\textwidth]{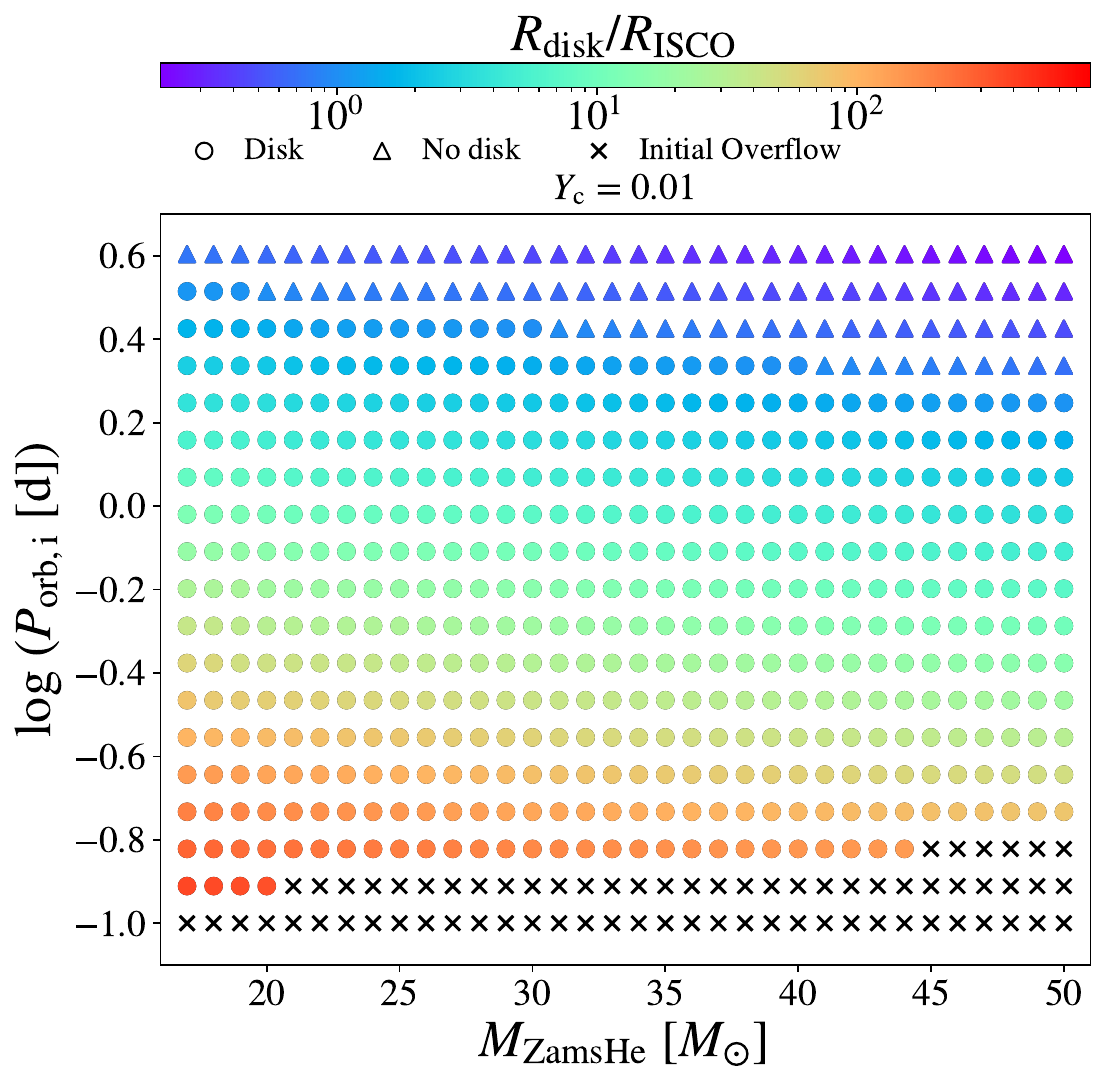}
     \caption{$R_{\rm disk}/R_{\rm \rm ISCO}$ at the central helium depletion as a function of the WR star mass at He-Zams. Dots correspond to the systems where an accretion disk can form, while triangles indicate systems without an accretion disk.}
     \label{Rdisk2} 
\end{figure}

In Figure~\ref{Rdisk1}, we show the ratio $R_{\rm disk}/R_{\rm \rm ISCO}$ for two representative BH-WR binaries with different initial orbital periods. We find that this ratio is higher for shorter orbital periods, reflecting its strong dependence on the orbital velocity (see Eq.~\ref{eq_criterion}). As the binary evolves, the ratio decreases with time. Nevertheless, it remains above unity when the WR star reaches central helium ($Y_{\rm c}$) depletion, indicating that an accretion disk can still form throughout the entire core-helium-burning phase. Furthermore, as shown by the extended grid of models in Figure~\ref{Rdisk2}, accretion disk formation is expected across the full WR mass range provided that the initial orbital period is shorter than approximately 2.0 d.

\label{lastpage}
\end{document}